%% file: main.tex
\documentclass[reprint,showpacs,amsmath,amssymb,xcolor=dvipsnames,prl,superscriptaddress]{revtex4-1}
\pdfoutput=1

\usepackage[pdftex]{graphicx}
\usepackage{bm}
\usepackage[pdftex,colorlinks=true,citecolor=blue,urlcolor=blue,linkcolor=black]{hyperref}
\usepackage{braket}
\usepackage{xcolor}
\usepackage[version-1-compatibility,binary-units,input-product=x,abbreviations=true]{siunitx}

\begin{document}


\title{Observing the emergence of a quantum phase transition - shell by shell} 
\author{Luca~Bayha}
	\email{bayha@physi.uni-heidelberg.de}
	\affiliation{Physikalisches Institut der Universit\"at Heidelberg, 69120 Heidelberg, Germany}
\author{Marvin~Holten}
   \email{mholten@physi.uni-heidelberg.de}
	\affiliation{Physikalisches Institut der Universit\"at Heidelberg, 69120 Heidelberg, Germany}
\author{Ralf~Klemt}
	\affiliation{Physikalisches Institut der Universit\"at Heidelberg, 69120 Heidelberg, Germany}
\author{Keerthan~Subramanian}
	\affiliation{Physikalisches Institut der Universit\"at Heidelberg, 69120 Heidelberg, Germany}
\author{Johannes Bjerlin}
    \affiliation{Mathematical Physics and NanoLund, LTH, Lund University, SE-22100 Lund, Sweden}
    \affiliation{The Niels Bohr Institute, University of Copenhagen, Blegdamsvej 17,DK-2100 Copenhagen \O, Denmark}
    \affiliation{Department of Physics and Astronomy, University of Southern California, Los Angeles, CA 90089-0484, USA}
\author{Stephanie~M.~Reimann}
    \affiliation{Mathematical Physics and NanoLund, LTH, Lund University, SE-22100 Lund, Sweden}
\author{Georg~M.~Bruun}
    \affiliation{Department of Physics and Astronomy, University of Aarhus, Ny Munkegade, DK-8000 Aarhus C, Denmark}
    \affiliation{Shenzhen Institute for Quantum Science and Engineering and Department of Physics, Southern University of Science and Technology, Shenzhen 518055, China}
\author{Philipp~M.~Preiss}
	\affiliation{Physikalisches Institut der Universit\"at Heidelberg, 69120 Heidelberg, Germany}
\author{Selim~Jochim}
	\affiliation{Physikalisches Institut der Universit\"at Heidelberg, 69120 Heidelberg, Germany}

\date{\today}

\maketitle


\textbf{
Many-body physics describes phenomena which cannot be understood looking at a systems' constituents alone \cite{Anderson1972}. Striking manifestations are broken symmetry, phase transitions, and collective excitations \cite{sachdev_2011}. 
Understanding how such collective behaviour emerges when assembling a system from individual particles has been a vision in atomic, nuclear, and solid-state physics for decades \cite{Bohr1975,Grebenev1998,Wenz2013,Launey2017}.
Here, we observe the few-body precursor of a quantum phase transition from a normal to a superfluid phase. The transition is signalled by the softening of the mode associated with amplitude vibrations of the order parameter, commonly referred to as a Higgs mode \cite{Bruun2014}.
We achieve exquisite control over ultracold fermions confined to two-dimensional harmonic potentials and prepare closed-shell configurations  of 2, 6 and 12 fermionic atoms in the ground state with high fidelity.
Spectroscopy is then performed on our mesoscopic system while tuning the pair energy from zero to being larger than the shell spacing. Using full atom counting statistics, we find the lowest resonance to consist of coherently excited pairs only. The distinct non-monotonic interaction dependence of this many-body excitation as well as comparison with numerical calculations allows us to identify it as the precursor of the Higgs mode.
Our atomic simulator opens new pathways to systematically unravel the emergence of collective phenomena and the thermodynamic limit particle by particle.}

\section{Introduction}
\begin{figure}
    \centering
	\includegraphics{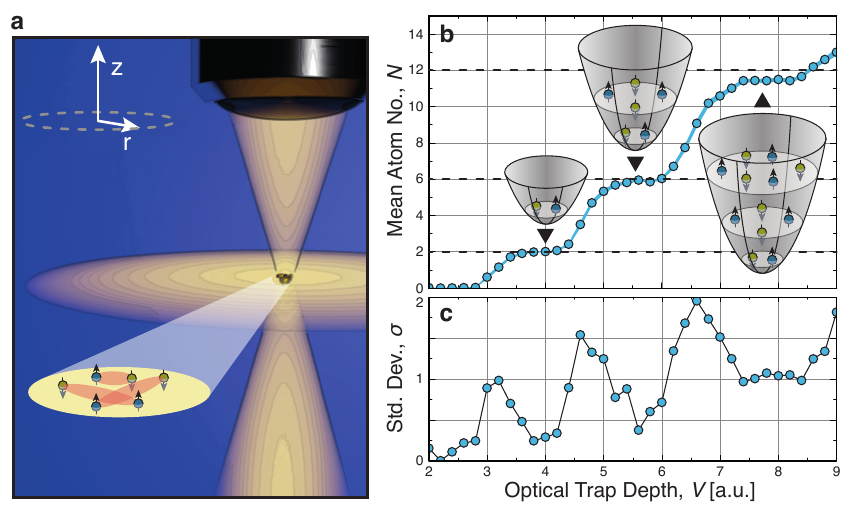}
    \caption{ \textbf{Deterministic preparation of two-dimensional     closed-shell configurations.} \textbf{a} Sketch of the experimental setup. The fermionic atoms are trapped in a single layer of an attractive optical lattice (horizontal disc) providing a tight confinement that freezes out motion along the z-direction. A superimposed tightly focused optical tweezer (vertical cone) provides radial harmonic confinement. \textbf{b} Mean trapped atom number as function of the final depth of the optical tweezer. 
    The stable atom numbers of 2,6 and 12 correspond to the closed-shell configurations of the two-dimensional harmonic oscillator. The third plateau lies slightly below the expected value due to the presence of holes in the third shell. The insets show sketches of the corresponding atom distribution. The standard error of the mean is indicated by the width of the band connecting the points.
    \textbf{c} Standard deviation of the detected atom number. At trap depths where the atom number reaches a plateau fluctuations are suppressed indicating the deterministic preparation of closed-shells of atoms.}
    \label{fig:main1}
\end{figure}
A key element of our understanding of nature is that macroscopic systems are characterized by the presence of phase transitions and collective modes, which cannot be extrapolated from the two-body solution \cite{Anderson1972}. While these effects in principle only exist in the thermodynamic limit, they are sometimes observed in surprisingly small systems. For instance, atomic nuclei consisting of only around 50 particles exhibit a collective mode spectrum consistent with a superfluid \cite{Bohr1975,Launey2017}. In liquid helium droplets superfluidity has been found to set in for similar particle numbers \cite{Grebenev1998}. Ultracold atoms offer the exciting possibility to study the onset of many-body physics in systems with full tunability of interactions, particle number, and single-particle spectra \cite{Bloch2008}.
The emergence of a Fermi sea was observed in a one-dimensional trap in Ref.\ \cite{Wenz2013}. Two- and three-dimensional systems promise even richer physics such as quantum phase transitions and symmetry breaking, as well as degenerate energy levels and single-particle spectra akin to the shell structure of atoms and nuclei.
\\In this work, we observe the few-body precursor of a quantum phase transition. The measurement relies on our experimental breakthrough in the preparation of a tunable number of fermionic atoms in the ground state of a two-dimensional (2D) harmonic potential. We study the interplay of the shell structure and Pauli blocking with the attractive interactions for closed-shell configurations. The competition of the gapped single-particle spectrum -- given by the confinement -- with the interactions gives rise to particular excitations exhibiting a non-trivial dependence on the attraction. We study this dependence and demonstrate that the modes consist of coherent excitations of particle pairs. Combined with a careful comparison to numerical calculations, this allows us to identify these excitations as few-body precursors of a Higgs mode associated with a quantum phase transition to a superfluid of Cooper pairs \cite{Bjerlin2016,som}. Comparing measured spectra for different atom numbers allows us to observe the approach towards the thermodynamic limit.
\\ In the many-body limit, a Higgs mode has been observed in cold atom, superconducting and ferromagnetic systems \cite{Sooryakumar1980,Ruegg2008,Bissbort2011,Endres2012,Matsunaga2013,Leonard2017,Katsumi2018,Behrle2018}. Our results pioneer the study of emergent quantum phase transitions and the associated Higgs mode starting from a few-body system.

\section{Experimental setup}
We perform our experiments with a balanced mixture of two hyperfine states of $^6$Li confined in a  trap created by the superposition of an optical tweezer (OT) and a single layer of an optical lattice (see Fig. \ref{fig:main1} a). The radial trapping frequencies of $f_\text{r} \approx 1000\, \text{Hz}$ are significantly smaller than the axial frequency $f_\text{z} \approx  6800 \, \text{Hz}$. Hence, for low temperatures and only a few occupied shells, the sample is in the quasi-2D regime and the dynamics along the third direction is frozen out. 
We prepare the ground state of up to 12 atoms in this trap by applying a novel spilling technique for quasi-2D systems based on the method in Ref.~\cite{Serwane2011}. A magnetic field gradient is applied and the power of the OT lowered such that only the lowest states inside the trap remain bound. This removes atoms initially occupying higher energy trap levels and the minimal power of the OT determines the initialized atom number. The degeneracy of the 2D harmonic oscillator is $k+1$ for the $k^\text{th}$ energy level resulting in the emergence of shells. The lowest three closed-shell configurations, i.e. where all states up to some energy are occupied and all other states are empty, contain 1, 3 and 6 fermions per spin state, respectively. Working with two hyperfine states, we expect closed-shell configurations of 2, 6 and 12 atoms. In the experiment, we observe plateaus for these 'magic' numbers as a function of the OT depth (Fig. \ref{fig:main1} b). The increased stability of the closed shells is also visible in the atom number fluctuations, which are strongly suppressed for closed-shell configurations (Fig. \ref{fig:main1} c).
An optimized sequence gives preparation fidelities of $97 \pm 2\, \%$, $93 \pm 3\, \%$   and $76 \pm 2 \, \%$, for 2, 6 and 12 atoms respectively. The deterministic loading of the ground state in two-dimensional systems is the first main result of our work and opens new prospects for quantum simulation with ultracold atoms. More details on the experimental procedure are provided in the Methods section \cite{som}.
\begin{figure}
    \centering
	\includegraphics{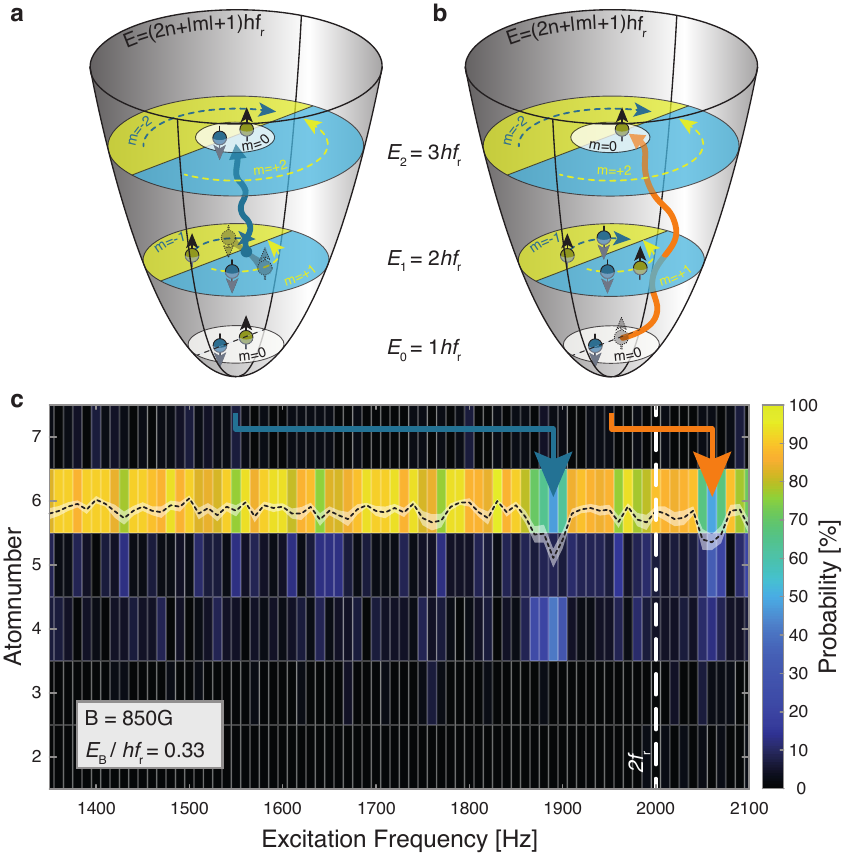}
    \caption{\textbf{Excitation spectrum for 6 particles.} Modulating the interaction strength at a frequency close to twice the radial trap frequency $f_\text{r}$ leads to excitations of pairs (a) and single particles (b). We start with 6 atoms in the ground state, modulate the interaction at different frequencies and count the number of atoms remaining in the lowest two shells. The  probability for detecting $N$ remaining atoms as a function of the modulation frequency is shown in (c). We observe the lowest resonance at 1890 Hz below twice the trap frequency, indicating mode softening. Remarkably, only the probability of detecting 4 atoms is increased, showing that this resonance consists of coherent superposition of pair excitations.
    The second resonance at 2060 Hz lies above twice the trap frequency consistent with a mean-field estimate. It is composed of single-particle excitations, as the probabilities for detecting both 4 and 5 atoms are enhanced.}
    \label{fig:main2}
\end{figure}
\\The closed-shell configurations are an interesting starting point to introduce interactions: While a gapless Fermi gas at zero temperature undergoes a phase transition from normal to superfluid at any attraction, a gap in the single-particle spectrum gives rise to a quantum phase transition from a normal to a superfluid phase at a certain critical interaction strength \cite{Kohomoto1990,Nozieres1999}. For the 2D harmonic oscillator, this means that for partly filled (open) shells this system will pair for arbitrarily weak attraction, whereas for completely filled (closed) shells and weak attraction the system is dominated by the energy gap to empty shells and pairing is suppressed.
Rich physics thus arises from the competition of interactions with the single-particle shell structure \cite{Heiselberg2002a,Bruun2002,Rontani2017,Bruun2014,Bjerlin2016}. 
Experimentally, we tune and control the interactions using a Feshbach resonance \cite{Zurn2013}.
Because the experiment is performed in a quasi-2D geometry there exists a two-body bound state for any attractive contact interaction. The binding energy  $E_\text{B}$ of the pair uniquely characterizes the interaction strength \cite{Randeria1990}.

\section{Excitation Spectra}
We utilize many-body spectroscopy to probe the effect of interactions on closed-shell configurations. The system is excited by modulating the axial confinement at frequencies far below the bandgap of the lattice, which only modulates the effective two-dimensional interaction strength between the different hyperfine states \cite{Idziaszek2006,som}. 
This interaction perturbation couples strongly to collective excitations driven by pairing correlations \cite{Bjerlin2016}. After modulation, all atoms excited to higher lying states are removed by a second spilling procedure and we count the remaining atoms. Repeating this procedure for different excitation frequencies gives the spectrum shown in Fig. \ref{fig:main2} (c).
\\The full counting statistics does not only contain the resonance positions, but also reveals the difference between pair and single-particle excitations.
\begin{figure*}[hbt!]
    \centering
	\includegraphics{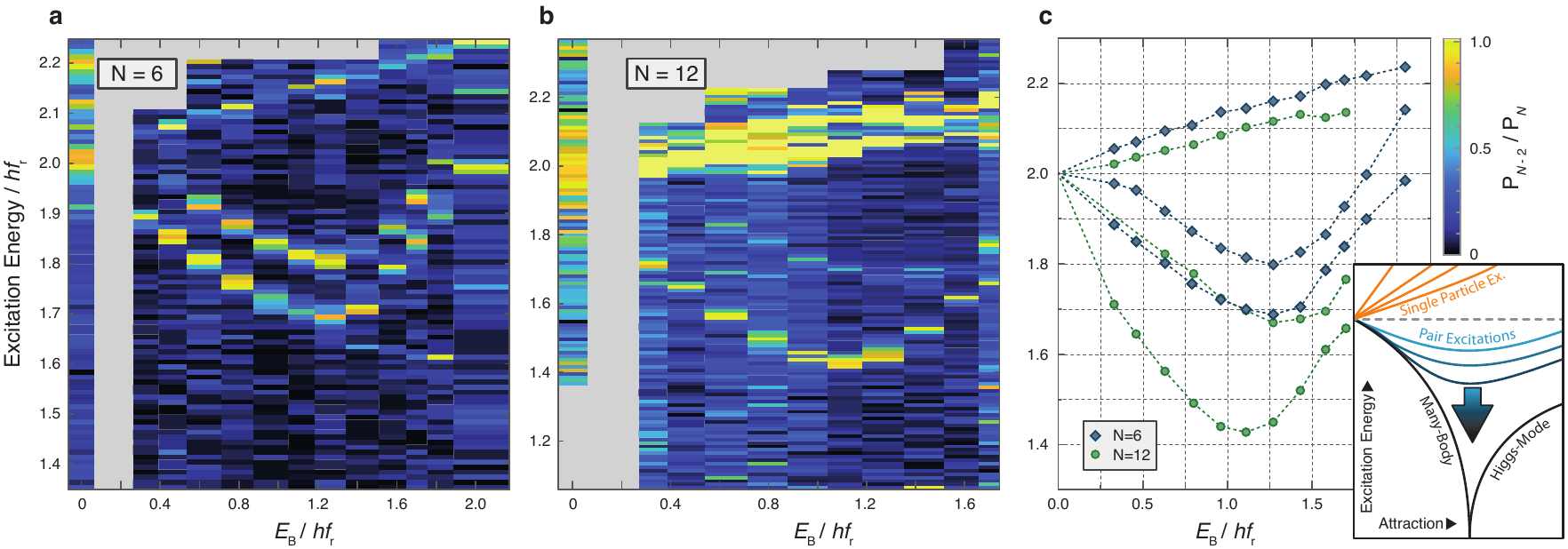}
    \caption{\textbf{Excitation spectra as function of interaction strength for 6 (a) and 12 (b) particles.} We show the probability of exciting a pair of particles $P_{N-2}$ divided by the probability of retaining the initial particle number $P_N$ normalized to the maximum of the lower pair excitation for each interaction strength. The system is excited by modulating the interaction strength for a fixed time for all interactions except for the spectra shown for $E_\text{B} = 0$, where the system is excited by modulating the radial trapping potential and where we show the normalized excitation probability $1 - P_N$. The resonance position for the lowest three excitations shown in (c) are extracted by Gaussian fits to the respective spectra. The error bars extracted from the fit are smaller than the symbol size. Comparing the data for 6 and 12 atoms we observe a deepening of the minimum of the pair excitation mode with particle number as expected for the transition to the many-body limit. Moreover, the position of the minimum  shifts to smaller interaction strength for more particles. The full counting statistics for (a) and (b) are shown in Extended Data Fig. \ref{fig:som6} and \ref{fig:som7}. The inset of (c) sketches the approach of the pair excitation (Higgs mode) in blue towards the many-body limit (black), upon increasing the particle number as discussed in Ref.  \cite{Bjerlin2016}. We scale the attraction by the critical attraction to compare to the thermodynamic limit.}
    \label{fig:main3}
\end{figure*}
The spectrum shows two resonances, where the probability of detecting 6 atoms is significantly reduced. The higher resonance at 2060 Hz lies above twice the noninteracting trap frequency  $ 2 f_\text{r}  \approx 2000\,  \text{Hz}$. 
This is consistent with an attractive mean-field potential from the atom cloud, which increases the effective trapping frequency. The mean-field energy is proportional to the atom density, thus predicting a larger interaction shift for the ground than for the more dilute excited state.
This higher resonance consists of single-particle excitations two shells up in energy. For the chosen drive strength and time there is a significant probability of exciting the system more than once. This explains the enhanced probabilities for detecting 4 and 5 atoms.
\\Mean-field theory however completely fails to explain the lower resonance at 1890 Hz, below twice the noninteracting trap frequency. Furthermore, the observed atom number distribution is strikingly different. Here only the probability of detecting 4 atoms is enhanced while all other probabilities are flat. Thus, at this frequency it is only possible to excite a \textit{single pair} of atoms and not individual atoms or two pairs. 
Both the energy and the atom number distribution are clear signatures of the collective nature of this excitation arising from the competition between the single-particle gap and the attractive interactions.
The spectrum is obtained for a binding energy of $E_\text{B} = 0.33 h f_\text{r}$, which is smaller than the single-particle gap and pairing is suppressed for closed shells due to Pauli blocking. However, by exciting a coherent superposition of particle pairs from the completely filled shell the remaining atoms can enhance their overlap by occupying the now empty states and thereby gain pairing energy. The excited particles form a pair in the otherwise empty shell and have a lower energy compared to two noninteracting particles in the same shell. Thus, the pair excitation lies below twice the trap frequency.\\
Next, we investigate the competition between pairing and the shell structure in more detail by tuning their relative strength using a Feshbach resonance.
The spectrum for different binding energies shown in Fig. \ref{fig:main3} (a) allows us to track the evolution of the different excitations  discussed above. The branch highest in energy shows a monotonous increase of frequency with interaction, as expected, from the increasing mean-field shift.
Remarkably, the lower two branches show a \textit{non-monotonic} behaviour. As we shall discuss in detail below, they correspond to coherent excitations of pairs with angular momentum $0$ and $\pm 2\,\hbar$.
For small interactions the energy of these excitations decreases with increasing attraction. This is due to the increasing gain in binding energy and the larger pair correlations in the excited state. This picture breaks down above an interaction strength of $ E_\text{B} \approx 1.1 h f_\text{r} $ where the lower mode energies start to increase with the attraction. In this regime the binding energy is comparable to the radial trap frequency and pairing becomes important also for the closed-shell ground state. Here, it is energetically favourable to have an admixture of higher lying harmonic oscillator levels to form a pair. Consequently, the ground state has significant pairing correlations and its energy decreases faster than that of the excited states. We identify the position of the minimal excitation gap  with the critical interaction strength.
\\To study the scaling of the spectrum towards the many-body limit, we fill one more shell in our trap, working with 12 particles. The corresponding excitation spectrum is shown in Fig. \ref{fig:main3} (b). Qualitatively the spectra for $N=12$ and 6 show the same features. For the larger system the number of states that are shifted upwards in energy above $2hf_\text{r}$  increases, rendering it impossible to resolve a single well-defined excitation peak. Importantly, the minima of the pair excitation branches below $2hf_\text{r}$ deepen and move to smaller interaction strengths for larger particle numbers, as evident form the resonance positions shown in Fig. \ref{fig:main3} (c). 
\\Crucially, the qualitative behaviour of this spectrum can be understood from many-body theory. In the thermodynamic limit, a closed-shell system undergoes a quantum phase transition from a normal to a superfluid phase with increasing attraction \cite{Bruun2001,Bruun2014}. As a generic feature of quantum phase transitions \cite{sachdev_2011}, this gives rise to a collective mode that goes soft at the transition point. In the case at hand, the lowest collective mode corresponds to the coherent excitations of time-reversed pairs across the gap. The energy cost of these excitations vanishes at the transition point reflecting that the system spontaneously forms Cooper pairs. From a broken symmetry perspective, the mode corresponds to amplitude vibrations in the order parameter (pairing strength) around its average value, which is zero/non-zero in the normal/superfluid phase. In the superfluid phase this mode is referred to as the Higgs mode. 
\\The pair excitation modes we observe in the experiment are the few-body precursors of the Higgs mode \cite{Bjerlin2016}: Due to the finite particle number, the phase transition is broadened to a crossover and the gap does not close completely. However, the lowest excitation corresponding to zero angular momentum retains the non-monotonic dependence on interactions and the pair correlation character. Adding more particles to the system decreases the minimal gap, consistent with an eventual complete gap closing in the many-body limit. When the Fermi energy increases, the relative importance of the single-particle gap decreases, and the minimal gap moves towards smaller binding energies. Both the softening of the mode and the shift to smaller critical binding energies when approaching the many-body limit are clearly visible when going from two to three closed shells. 
This interpretation is explicitly confirmed by comparing to a numerical diagonalisation of the microscopic Hamiltonian \cite{som}. The higher non-monotonic branch in fact corresponds to two nearly degenerate modes consisting of coherent excitations of pairs with angular momentum $\pm 2\,\hbar$. They are precursors of Higgs modes of a superfluid with higher angular momentum Cooper pairs. Although modulating the interaction strength does not add angular momentum, these modes are visible due to a slight breaking of the circular symmetry of the trap \cite{som}.

\section{Coherent Drive}
At all interaction strengths shown in Fig. \ref{fig:main3} the Higgs mode is a well-defined excitation: The line width is on the order of $10 \, \text{Hz}$ and thus much smaller than the excitation energy. 
To  probe the stability of the excited state we drive the 6 particle system for variable times at the frequency of the lower pair excitation mode. We observe oscillations between the probability of detecting 4 and 6 particles in the lowest two shells indicating the coherent formation and destruction of a pair in the excited shell at a Rabi rate of $8.0 \pm 0.1 \, \text{Hz}$. The 1/e decay rate of the oscillation of $4.5 \pm 0.5\, \text{Hz}$ gives a quantitative upper limit on the lifetime of the Higgs mode, exceeding the transition frequency of 1480 Hz by a factor of more than 300. 
This long lifetime of the excited state can be attributed to the discrete level spectrum of our trap, limiting the possible decay channels \cite{Bruun2014}.
\begin{figure}
    \centering
	\includegraphics{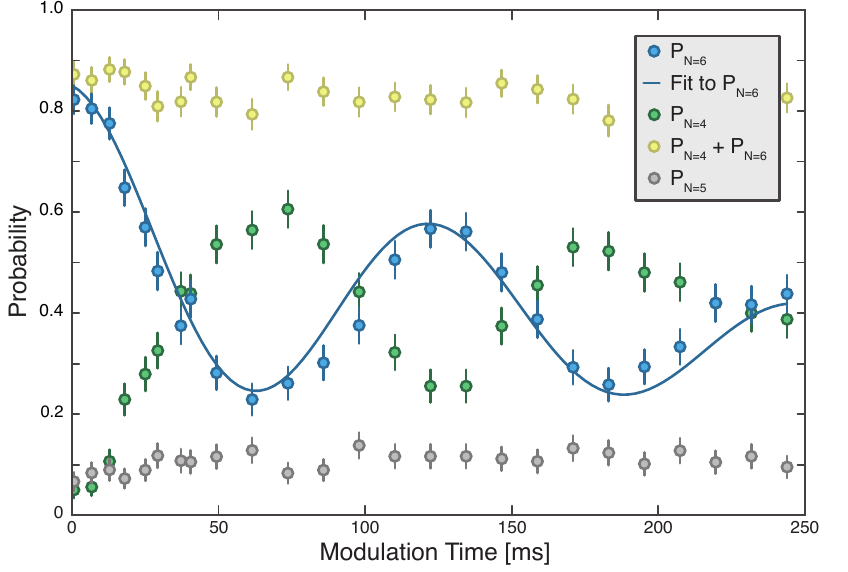}
    \caption{ \textbf{Coherent driving of the lower Higgs mode.} The interactions are modulated at the resonance frequency of the lower Higgs mode for variable times. The probabilities for  4 and 6 particles coherently oscillate out of phase showing the pair character of the excitation. Coupling to other states is negligible as seen from the almost constant value of  $P_{N=4} + P_{N=6}$ and the fact that $P_{N=4}$ and $P_{N=6}$ converge to the same value. Thus, the ground state plus the lower Higgs mode can be described as a coherently driven two-state system. We fit $P_{N=6}$ with an exponentially damped Rabi oscillation. The data is taken for  $E_\text{B}  = 0.57 h f_\text{r}$. The error bars represent the standard error.}
    \label{fig:main4}
\end{figure}
\section{Conclusion and Outlook}
In conclusion, we have shown that systems consisting of only a few particles exhibit precursors of a quantum phase transition to a superfluid phase with an associated Higgs mode present in the thermodynamic limit. 
In addition to the emergence of pairing, the achieved degree of control over this mesoscopic system will allow us to study thermalisation in isolated quantum systems \cite{DAlessio2016} and fermionic superfluidity at its fundamental level.
As a next step, we will to go beyond the excitation spectrum studied here and investigate the emergence of pair correlations across the precursor of the normal- to superfluid transition directly in momentum space. Scaling up the system will allow us to  observe the emergence of Cooper pairs and the Goldstone mode, which requires the pair size to be much smaller than the system size \cite{Bruun2001}.
\paragraph*{Data availability}
The data that support the findings of this study are available from the
corresponding authors upon reasonable request.

\paragraph*{Acknowledgements}
The experimental work has been supported by the ERC consolidator grant 725636, the Heidelberg Center for Quantum Dynamics,  the DFG Collaborative Research Centre SFB 1225 (ISOQUANT) and the European Union’s Horizon 2020 research and innovation program under grant agreement No.~817482 PASQuanS. P.M.P acknowledges funding from the Daimler and Benz Foundation.
S.M.R. an J.B. acknowledge financial supported by the Swedish Research Council and the Knut and Alice Wallenberg Foundation. G.M.B acknowledges financial support from the Independent Research Fund Denmark - Natural Sciences via Grant No.~DFF~-~8021-~00233B.

\paragraph*{Author Contributions}
L.B.\ and M.H.\ contributed equally to this work. L.B.,\ M.H.,\ and K.S.\ performed the measurements and analyzed the data. J.B.,\ S.M.R.,\ and G.M.B.\ developed the theoretical framework. J.B. performed the numerical calculations. P.M.P.\ and S.J.\ supervised the experimental part of the project. All authors contributed to the discussion of the results and the writing of the manuscript.

\paragraph*{Competing Interest}
The authors declare no competing interests.

\paragraph*{Correspondence and requests for materials}
should be addressed to L.B.\ and M.H.

\bibliographystyle{naturemag.bst}
\bibliography{Higgs}

\cleardoublepage
\setcounter{figure}{0}
\renewcommand{\figurename}{Extended Data Figure}

\newpage

\section*{Methods}
\input{supplement_v2}

\newpage

\newpage
\input{arxivExtendedFigures}
\end{document}

%% file: supplement_v2.tex
\paragraph*{\textbf{Experimental sequence}}
The experimental sequence starts by transferring a gas of $^6$Li atoms from a magneto-optical trap (MOT) into a red-detuned crossed-beam optical dipole trap. Here, we make use of radio frequency pulse sequences to prepare a balanced mixture of the two  hyperfine states $\ket{1}$ and $\ket{3}$ of the ground state of $^6$Li. We label the hyperfine states according to their energy from lowest ($\ket{1}$) to highest ($\ket{6}$).

After a first evaporative cooling stage in the crossed-beam optical dipole trap we transfer approximately 1000 atoms into a tightly focused optical tweezer (OT). In the OT quantum degeneracy is reached by spilling to around 20 atoms close to the ground state of the approximately harmonic confinement by the procedure described in Ref.\ \cite{Serwane2011}. Subsequently, we begin the crossover to a quasi-2D system (see Extended Data Fig. \ref{fig:som1}). This is achieved with an adiabatic transfer of the atoms from the effective confinement of the OT alone with $f_\text{r}: f_\text{z} \approx 5:1$ to $f_\text{r}: f_\text{z} \approx 1:7$ in a combined potential of OT and a single layer of a  one-dimensional optical lattice. To this end, we lower the radial trap frequency $f_\text{r}$ of the OT  from approximately $20\,\text{kHz}$ to $1\,\text{kHz}$. This is done by ramping the pattern displayed on a Spatial Light Modulator in $20\,\text{ms}$, which changes the aperture of the optical setup creating the OT. This changes the waist of the OT from $\approx 1 \, \mu \text{m}$ to $\approx 5 \, \mu \text{m}$. The transfer is performed at a magnetic field of 750 G to have sizeable coupling between the different states. The axial confinement is solely defined by the optical lattice with $f_\text{z}\approx 6.8\,\text{kHz}$. The measurements with 6 atoms were performed at a final radial trap frequency of $f_\text{r} = 1001\,\text{Hz}$ while the 12 atom data were taken at $f_\text{r} = 992\,\text{Hz}$.
\\In the combined trap we create closed-shell configurations of the quasi-2D harmonic oscillator, by applying a magnetic gradient of approximately $70\, \text{G/cm}$ in axial direction and reducing the power of the OT such that only the lowest one to three energy shells remain bound. The spilling procedure is performed at 750G, where the interaction energy is sufficiently small such that one recovers the noninteracting shell structure. After preparation, we increase the OT power back up until we recover the trap frequencies and aspect ratio discussed in the main text. 

The measurements of the excitation spectrum are performed by a sinusoidal modulation of the power of either the OT or of the optical lattice with frequency $f_\text{ex}$ for $t=400\,\text{ms}$. To detect excitations we implement a final spilling stage after which we transfer all the remaining trapped atoms from the OT back into the MOT. In the MOT we are able to determine the total atom number in both spin states with a fidelity exceeding $99\,\%$. The latter is achieved by integrating the total fluorescence signal of the MOT for one second on a CCD camera. Excitations of the system show up by a reduced atom number compared to the closed-shell ground states of two, six or twelve atoms.

\paragraph*{\textbf{Experimental Parameters}}

To compute the binding energy $E_\text{B}$, we use the exact analytical solution of the Schr\"odinger equation for two ultracold atoms confined in an axially symmetric harmonic potential provided by \cite{Idziaszek2006}. The solution depends on three parameters: the 3D s-wave scattering length $a_\text{3D}$ and the trap frequencies in radial ($f_\text{r}$) and axial ($f_\text{z}$) direction. The scattering length is set via the magnetic offset field $B$ by making use of the Feshbach resonance of the $\ket{1}$ - $\ket{3}$ mixture at $B_\text{0}=690\,\text{G}$ \cite{Zurn2013}. To verify the accuracy of the calculated value for the binding energy $E_\text{B}$, we compare the analytical result for the excitation energy of the two-body problem to the spectrum measured for two atoms (see Extended Data Fig. \ref{fig:som8}).
\\The trap frequencies are determined using the same sequence as explained above. The only difference is that the system is excited by modulating the confinement at a magnetic field of  $B=568\,\text{G}$, i.e. at the zero crossing of the scattering length.
In the noninteracting system the lowest monopole excitation is at $f=2f_\text{r} (2f_\text{z})$, for radial (axial) excitations. This allows us to extract the frequencies by modulating the harmonic confinement in the respective directions. The measurements for $f_\text{r}$ for six and twelve atoms are shown at $E_\text{B} = 0$ in Fig. \ref{fig:main3} (a) and (b) of the main text. For the axial direction we find $f_\text{z} = 6803\pm2 \,\text{Hz}$.

\paragraph*{\textbf{Different Modulation Schemes}} 
As discussed in the previous section, there are two different schemes that we use to drive excitations above the closed-shell ground state. All the data shown in the main text, except the spectra taken at $E_\text{B} = 0$, are recorded by modulating  the  optical lattice. Since, we modulate the depth of the axial confinement  at frequencies well below its band gap, this does not create excitations along this tightly confined direction. The wave function adiabatically follows the potential change and is compressed periodically. The effective 2D interaction is obtained by integrating out the wave function along the third direction. Thus this  effectively only modulates the two-dimensional binding energy. The strength of the modulation corresponds to a change of $E_\text{B}$ by approximately 2\%.
\\For reference, we compare the modulation of the radial trapping potential with the modulation of interactions at 300G. We find that both schemes lead to different relative transition probabilities for the Higgs and the other excited modes (see Extended Data Fig. \ref{fig:som2}). The locations of the respective excitations in the spectrum remain unaffected. The qualitative result that a modulation of interaction strength leads to an increased transition matrix element of the Higgs mode is consistent with the few-body calculation by Bjerlin \textit{et al.\ }\cite{Bjerlin2016}. Consequently, $f_{z}$-modulation has been applied for all the data shown in the main text, except for the spectra taken at $E_\text{B} = 0$.

\paragraph*{\textbf{Anisotropy and Anharmonicity}}
The small size of our OT results in a finite anharmonicity of the trapping potential. The transition frequency from the lowest shell two shells up is $\sim 10 \%$ larger than the transition frequency from the second shell to the fourth shell. The  anharmonicity extracted from the noninteracting spectra matches our expectation due to the finite size of the optical tweezer. Its waist of $\sim 5\, \mu \text{m}$ is of the same magnitude as the harmonic oscillator length $ l_{\text{ho}} = \sqrt{\hbar / \omega  m } \approx  1.3\, \mu \text{m}$ and the atoms probe the non-harmonic parts of the trap.
In addition,  the trap shows a slight anisotropy $\left(\omega_\text{x}-\omega_\text{y}\right)/\left(\omega_\text{x}+\omega_\text{y}\right)$ of approximately $2\,\%$. These corrections should not affect the qualitative behaviour of the measured spectra for the interacting system. However, they might quantitatively change the coupling strengths to the different modes and the exact shape of the Higgs mode. We neglect the influence of the anharmonicity for calculating the binding energy. A comparison of the calculated and measured excitation energies for two particles shown in Extended Data Fig. \ref{fig:som8}, confirms that this is only a small effect, as expected.

\paragraph*{\textbf{Numerical Modeling}}
We model the experiment using a trapping potential of the form

\begin{align}
V=V_\mathrm {2D}(x,y) + \frac12m\omega_z^2z^2.
\label{Potential}
\end{align}
Here, $V_\mathrm {2D}$ describes the potential in the $xy$-plane, which is provided by the Gaussian profile of the OT so that

\begin{align}
    V_\mathrm {2D}(x,y)= \frac{A}{2}\hbar\omega_r\cdot [1 - e^{-(\gamma x^2  +\frac{ y^2}{\gamma})/( l_{\text{ho}}^2A)} ],
    \label{eq:2Dpotential}
\end{align}
with the (lowest order) harmonic trapping frequency $\omega_r$.
The parameter $\gamma$ controls the ratio of the trap frequencies in the x- and y-directions and hence the anisotropy. The parameter $A$ is the depth of the trap (in units of $\hbar \omega_r$) and determines the anharmonicity.

Pure contact interactions in 3D are represented by a term $g_{3D}\delta(\mathbf{r}_k-\mathbf{r}_l)$. We assume $\omega_z\gg \omega_r$ so that that the fermions reside in the lowest harmonic oscillator state along the $z$-direction. Consequentially, the $z$-direction can be integrated out yielding a quasi-2D model with an effective coupling strength $\widetilde g$~\cite{Bloch2008}. We end up with the effective 2D $N$-particle Hamiltonian

\begin{equation}
    \hat{H}_\mathrm{2D}= \sum^N_{i=1}[ \frac{ -\hbar^2  }{2m} \mathbf{\nabla}_i^2+ {V}_\mathrm{2D}({\mathbf r}_i )] 
    + {\widetilde g} \sum_{kl} \delta  \left( \mathbf{r}_k - \mathbf{r}_l \right), \label{eq:Hamiltonian}
\end{equation}
where  $\mathbf{r}_i=\left(  x_i,y_i \right)$ is the coordinate of  particle  $i$,  $\mathbf{\nabla}^2_i=\partial^2_{x_i} + \partial^2_{y_i}$, and  $k$ and $l$ denote fermions in the two different hyperfine (pseudo spin) states. Only fermions with different spin interact for a contact potential.

We solve the Hamiltonian numerically by using a single-particle basis of 2D isotropic harmonic oscillator states $\ket{m,n}$ with energies $(2n + |m|+ 1)\hbar \omega_r $. They are determined by principal and angular momentum quantum numbers  $n=0,1,2...$ and $m =0, \pm 1, \pm 2...$ respectively. 
Following Ref.~\cite{Bjerlin2016}, we employ a two-parameter cutoff scheme in which we define a single-particle subspace by the highest allowed total single-particle energy $E_\mathrm{sp}^\mathrm{cut}$, and then define the relevant many-body subspace in terms of the maximum many-body energy.

For circular symmetry ($\gamma=1$) the many-body eigenstates split into subspaces with different total angular momentum  $L_z$. In this case, it is sufficient to consider the $L_z=0$ subspace only, since a modulation of the trapping frequency $\omega_z$ does not impart angular momentum onto the ground state. However, the experimental trap is slightly anisotropic ($\gamma\approx 0.99$) and as a result a finite coupling between subspaces with different $L_z$ has to be considered.
In order to accelerate the convergence of our calculations for an approximately circular trap, we neglect negative angular momentum states in our basis and include a subspace of total angular momenta $L_z =0, \hbar$ and $2\hbar$ only.  This enables us to use a higher energy cutoff by exploiting the (approximate) symmetry between the  positive and negative $L_z$ states.

Care has to be taken when treating the $\delta(\mathbf r)$-interaction in Eq.~\eqref{eq:Hamiltonian}, which causes an ultraviolet divergence. 
This divergence can be regularized by expressing the coupling constant $\widetilde g$ and the cut-off in terms of the two-body binding energy $E_\text{B}$ ~\cite{Bjerlin2016,Rontani2017}. The parameters of the 2D potential (Eq.~\ref{eq:2Dpotential}) used in the numerical calculations are obtained by comparing to the lowest monopole excitation observed experimentally with six noninteracting particles. This yields $\gamma\approx 0.99$ and $A=20$ for the anisotropy and anharmonicity respectively.

The combination of a slightly broken circular symmetry  with the  high degree of collectivity when passing through a few-body precursor of a quantum critical point, makes the modelling by a full numerical diagonalization of the many-particle Hamiltonian a  highly non-trivial task. The complexity of configuration-interaction diagonalization methods often makes it challenging to reach full convergence in terms of the used basis set, which we also found to be  the case in our present simulations. We employed a parallel full diagonalization using implicitly restarted Arnoldi routines for sparse matrices of bases with up to $\sim 10$ million states. Trends going towards larger numbers of basis states were carefully analyzed. In this way, we were able to reach consistent results for the $N=6$ particle system, whereas reasonably converged numerical calculations for a $N=12$ particle system in a realistic trap are currently beyond our reach. We emphasise, however, that the qualitative  features of the spectra including the presence of three Higgs-like modes with a non-monotonic behaviour are reassuringly robust with increasing basis set. In general, we found that the essential features of the Higgs modes, such as the minimum in energy and the strong coupling via interaction modulation, become more pronounced with increasing basis set size.

\paragraph*{\textbf{Comparing Experimental and Numerical Spectra}}
Extended Data Fig.~\ref{fig:som10} (a) shows the calculated as well as experimentally observed spectrum for the $N=6$ particle system, with  $A=20$ and $\gamma=0.99$. The numerical calculcation includes states up to $E^{\text{cut}}_{\text{sp}}=10 \hbar \omega_\text{r}$ and up to a many-body energy of $28 \hbar\omega_\text{r}$. We see that there is reasonable agreement between theory and experiment and that all qualitative features in the spectrum are recovered by the calculations. In particular, the existence of two non-monotonic Higgs branches is confirmed by the calculations. 

The lowest branch connects smoothly to the $L_z=0$ Higgs mode for the isotropic case, whereas the two higher modes connect smoothly to the $L_z=\pm2\hbar$ Higgs modes when  the anisotropy goes to to zero ($\gamma=1$). The latter modes have a higher energy because they describe Cooper pairing with finite angular momentum, and are almost degenerate due to the small anisotropy. Since our numerics only include positive angular momentum states, only one of the $L_z=\pm2\hbar$ Higgs modes is visible in the calculated spectrum. In the experimental data they appear as a single resonance due to the small energy splitting for small anisotropy.

Extended Data Fig.~\ref{fig:som10} (b) shows the spectrum weighted with the matrix element 
\begin{align}\label{eq:matrix}
    \Gamma^E_{int}=|\langle G | \sum_{k,l} \delta (\mathbf{r}_k - \mathbf{r}_l) |E\rangle |^2,
\end{align}
which gives the coupling between the ground  $|G\rangle$ and excited state $|E\rangle$ when the interaction strength is modulated in the experiment. The slight breaking of the circular symmetry leads to the coupling of the ground state to all  three Higgs modes.
This plot also highlights why the manifold of states around the energy $\sim\hbar\omega_r$ is not observed experimentally: These states correspond to exciting one fermion one shell up, which changes the angular momentum  by  $\pm\hbar$. The trap anisotropy leads to quadrupole excitations solely, where the angular momentum changes by $\Delta L_z = \pm 2 \hbar $. As a result, these modes are not visible in the experimental spectrum.

\paragraph*{\textbf{Limitations of the Model}}
We note that the agreement between the theoretical and experimental spectra becomes worse in the region where the binding energy is significantly larger than the critical binding energy. There are two main reasons that explain this behaviour. First, this region corresponds to the few-body precursor of the BEC regime of dimers and the large binding energy requires a cut-off beyond what is numerically feasible. Second, the modelled potential only approximates the actual experimental confinement. The fitting of the potential parameters is only performed for a small set of experimental values, all within the lowest few harmonic shells. This allows for a qualitative simulation of the low-lying excitation spectra observed in the few-body experiments but we did not perform a systematic fit including all observed modes. The reason is that the experimental  trap is not precisely described by Eq.~\eqref{eq:2Dpotential}. There are deviations away from the Gaussian profile especially close to the continuum. More importantly, the experimental aspect ratio $\omega_z/\omega_r\simeq6.8$ implies that 3D effects become important at higher energies. These effects are not included in  our model making a more quantitative fitting procedure less meaningful.

In conclusion, the agreement between the  numerical calculations and the experiment confirms the physical interpretation of the data. In particular, we are indeed observing a few-body precursor of a quantum phase transition with the associated emergence of a Higgs mode. We note that obtaining an even better quantitative agreement requires a more accurate determination of the shape of the trap, inclusion of 3D physics for higher energies, and the use of substantially larger computer resources. All qualitative features of the low energy spectrum, however, were found to be insensitive towards these effects.

%% file: arxivExtendedFigures.tex
\newpage
\begin{figure}[h]
    \centering
	\includegraphics{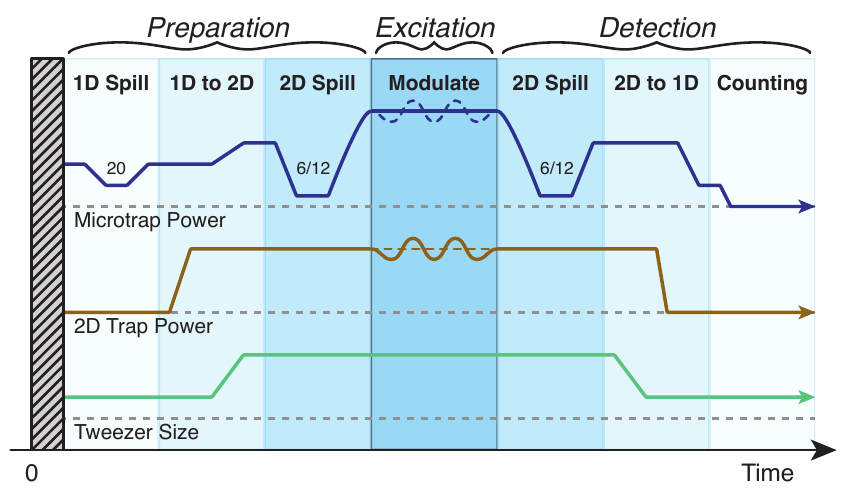}
    \caption{ \textbf{Experimental protocol.} The sequence can be separated into three parts. First, several evaporation and spilling stages are combined with a transfer from a quasi-1D to a quasi-2D trap geometry. This is needed to prepare closed-shell ground state configurations of up to 12 atoms. Next, we excite the system at some defined frequency $f_\text{ex}$ and magnetic offset field $B$ using a sinusoidal modulation of either the radial or axial confinement. Detection is implemented by spilling to the ground state a second time and a transfer of all remaining atoms to the MOT where we count their number.}
    \label{fig:som1}
\end{figure}

\begin{figure}
    \centering
	\includegraphics{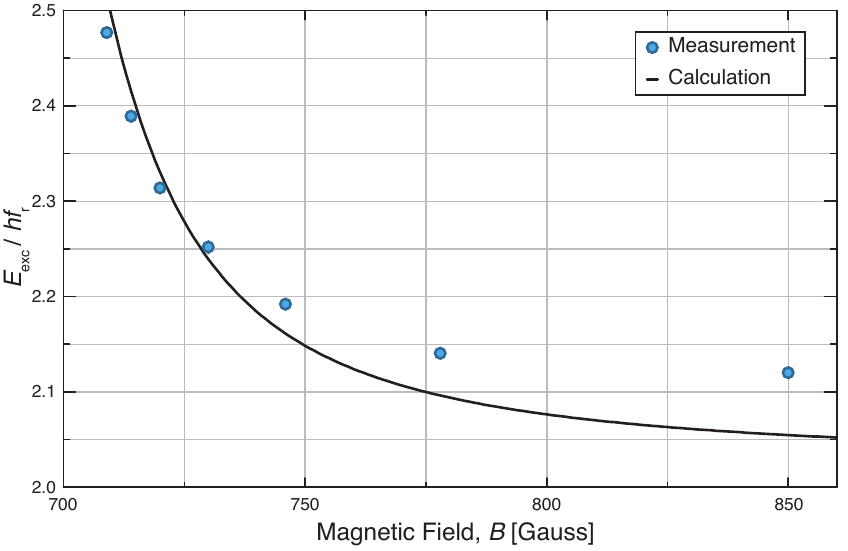}
    \caption{ \textbf{Excitation spectrum for two particles.} We define the two-body excitation energy $E_\text{exc}$ as the energy difference between the ground state and the lowest monopole excitation of the two atom system. It is measured using the same modulation scheme as for the Higgs mode. The system is initialized with one filled shell, i.e. two particles. The analytical solution of the two-body problem (solid line) shows good agreement with the measurement (blue points). Error bars are extracted from the fit to the spectrum and are smaller than the data points.}
    \label{fig:som8}
\end{figure}

\begin{figure}
    \centering
	\includegraphics{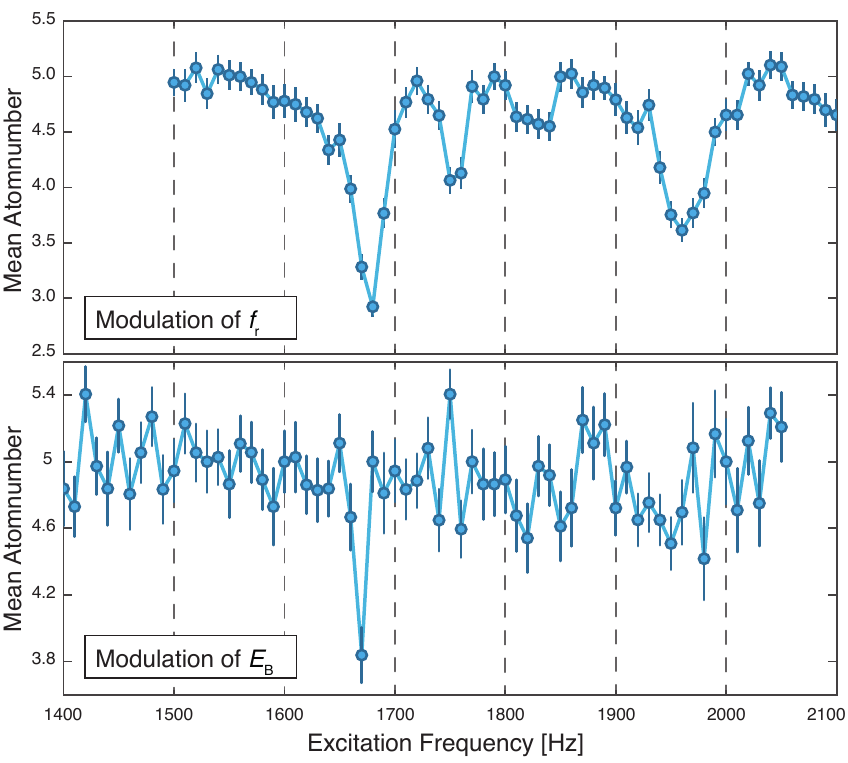}
    \caption{ \textbf{Comparison of different modulation schemes.} Modulating the trap frequency in radial direction $f_\text{r}$ leads to higher transition probabilities for the higher excited states (top) compared with modulation of the axial confinement $f_\text{z}$(bottom). Modulating the axial confinement effectively only modulates the interaction strength, which couples predominantly to the pair excitation mode. The data is taken for $ E_\text{B}=0.09 \, h f_\text{r}$. For this measurement the radial trap frequency was $2f_\text{r}= 1660 \,\text{Hz}$.  Error bars show the standard error of the mean.}
    \label{fig:som2}
\end{figure}

\begin{figure*}
    \centering
	\includegraphics{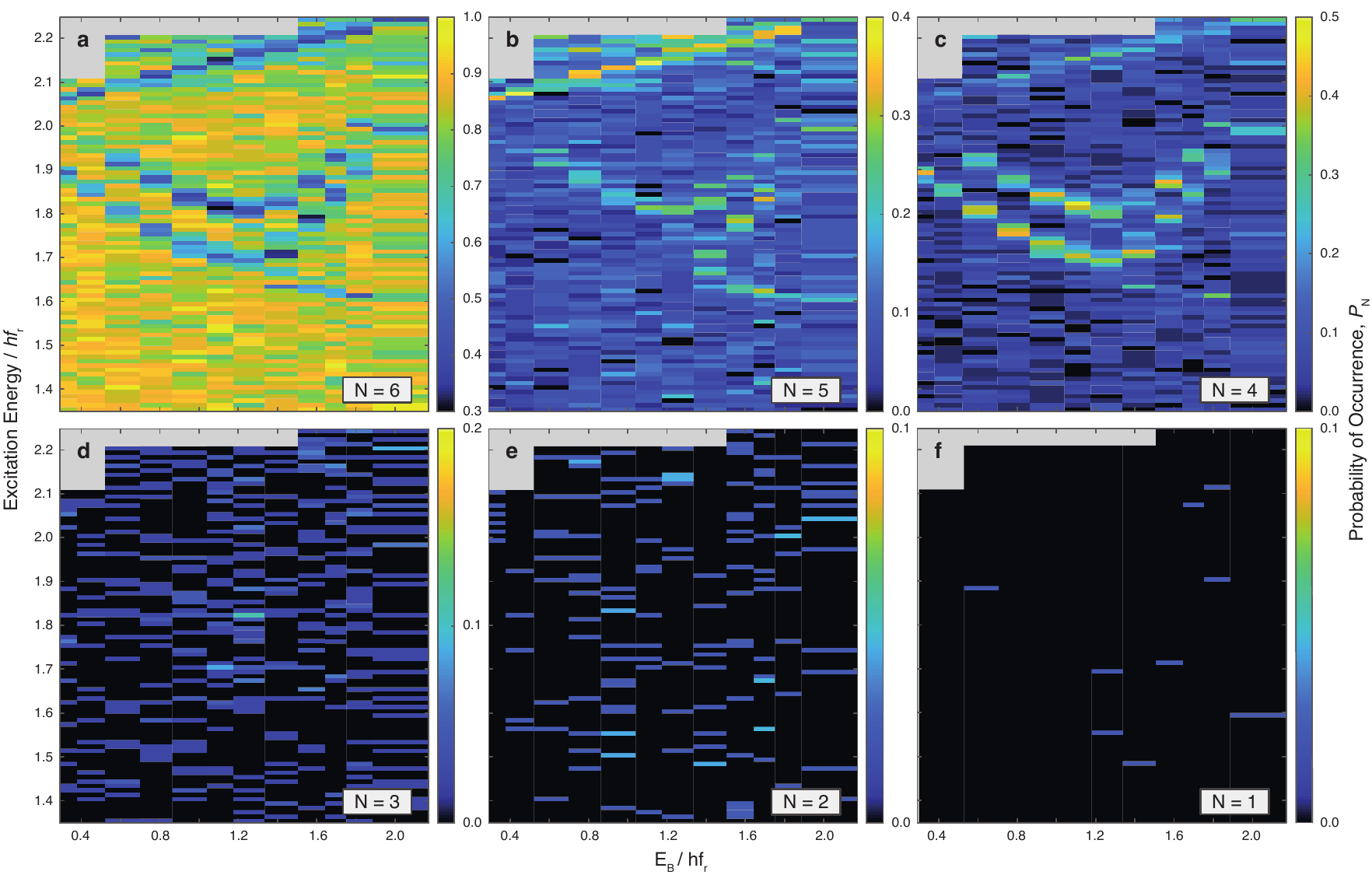}
    \caption{ \textbf{Probabilities of different  atom numbers remaining in the lowest 2 shells for the $N=6$ initial state.} The different panels a-f show the probabilities for different remaining atom numbers after modulating the 6 atom ground state with a defined frequency given on the y axis and subsequent removal of excited atoms. All possible excitations manifest themselves by a reduced probability of remaining in the ground state of 6 atoms (a). We find that the lowest excitation, or Higgs mode, mostly consists of excitations to four atoms (c), while the higher excited peaks are predominately generated by the loss of a single atom (b).}
    \label{fig:som6}
\end{figure*}
\begin{figure*}
    \centering
	\includegraphics{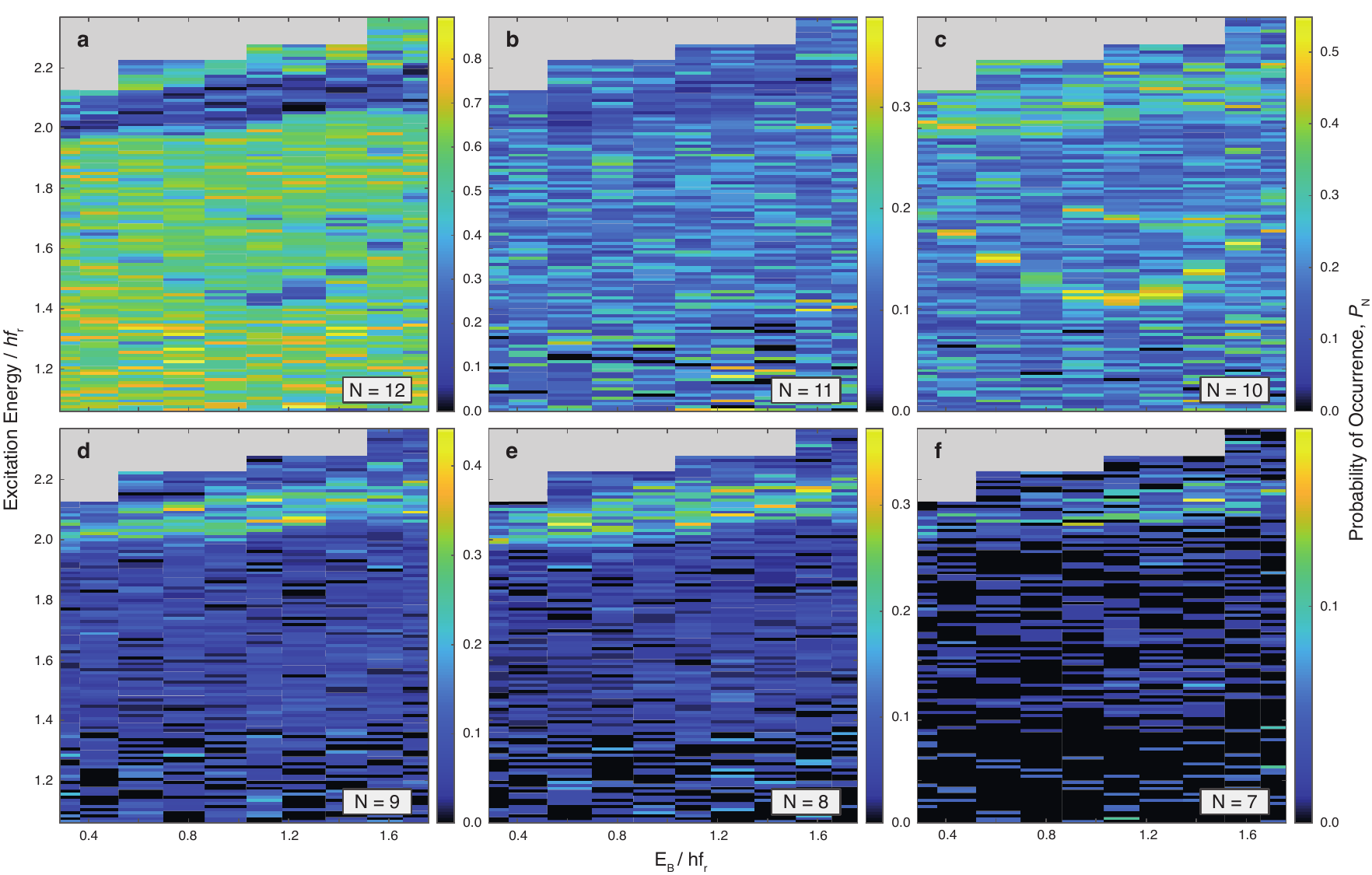}
    \caption{ \textbf{Probabilities of different atom numbers remaining in the lowest 3 shells for the $N=12$ initial state.} The different panels a-f show the probabilities for different remaining atom numbers after modulating the 12 atom ground state with a defined frequency given on the y axis and subsequent removal of excited atoms. All possible excitations manifest themselves by a reduced probability of remaining in the ground state of 12 atoms (a). We find that lowest excitation, or Higgs mode, mainly consists of excitations to ten atoms (c), while the higher excited peaks are predominately generated by the loss of even more atoms (d-f).}
    \label{fig:som7}
\end{figure*}

\begin{figure*}
    \centering
	\includegraphics{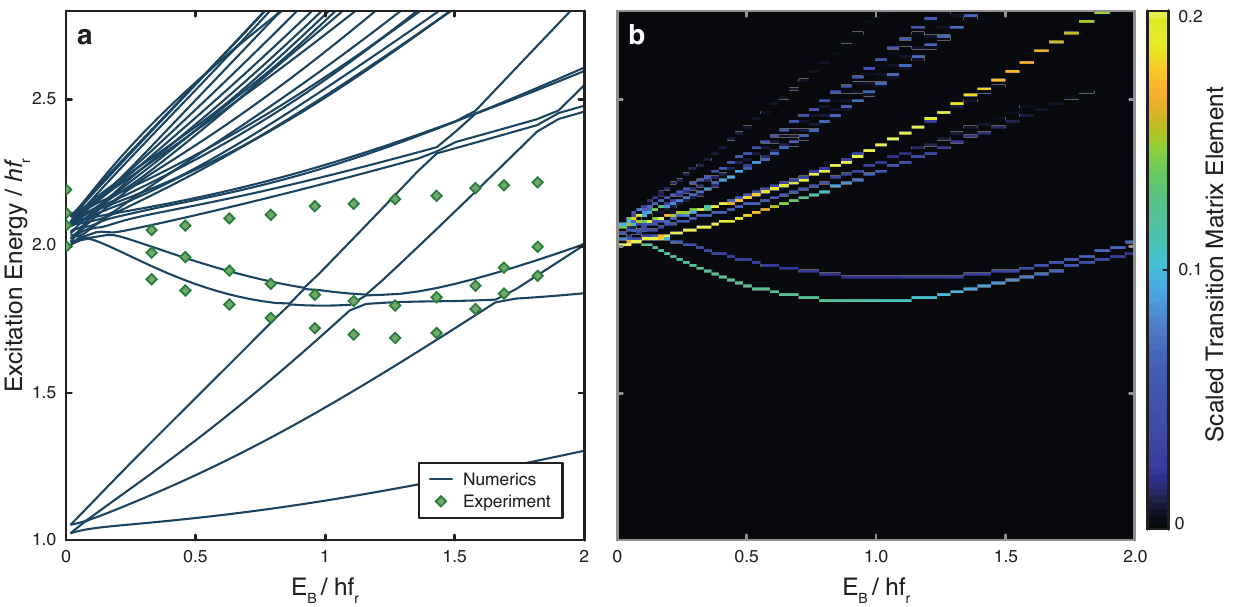}
    \caption{ \textbf{Numerically calculated excitation spectrum for 6 particles.} Panel (a) shows the level spectrum obtained by exact diagonalisation with parameters $A=20$ and $\gamma = 0.99$ for the potential as well as the experimental results. The calculation includes states up to $E^{\text{cut}}_{\text{sp}}=10 \hbar \omega_\text{r}$ and up to a many-body energy of $28 \hbar\omega_\text{r}$. For comparison the experimental data is shown by green diamonds. Values and errors bars are obtained as in Fig. \ref{fig:main3} c. In (b) the numerically calculated excitation spectrum for a modulation of the interaction strength is shown. As in the experiment we observe that this modulation couples to two non-monotonous modes. Note that the calculations performed for (b) employ a smaller cut-off than in (a) due to the computational demand in calculating the matrix element (Eq. \ref{eq:matrix}).}
    \label{fig:som10}
\end{figure*}

%% file: main.bbl
\begin{thebibliography}{29}%
\makeatletter
\providecommand \@ifxundefined [1]{%
 \@ifx{#1\undefined}
}%
\providecommand \@ifnum [1]{%
 \ifnum #1\expandafter \@firstoftwo
 \else \expandafter \@secondoftwo
 \fi
}%
\providecommand \@ifx [1]{%
 \ifx #1\expandafter \@firstoftwo
 \else \expandafter \@secondoftwo
 \fi
}%
\providecommand \natexlab [1]{#1}%
\providecommand \enquote  [1]{``#1''}%
\providecommand \bibnamefont  [1]{#1}%
\providecommand \bibfnamefont [1]{#1}%
\providecommand \citenamefont [1]{#1}%
\providecommand \href@noop [0]{\@secondoftwo}%
\providecommand \href [0]{\begingroup \@sanitize@url \@href}%
\providecommand \@href[1]{\@@startlink{#1}\@@href}%
\providecommand \@@href[1]{\endgroup#1\@@endlink}%
\providecommand \@sanitize@url [0]{\catcode `\\12\catcode `\$12\catcode
  `\&12\catcode `\#12\catcode `\^12\catcode `\_12\catcode `\%12\relax}%
\providecommand \@@startlink[1]{}%
\providecommand \@@endlink[0]{}%
\providecommand \url  [0]{\begingroup\@sanitize@url \@url }%
\providecommand \@url [1]{\endgroup\@href {#1}{\urlprefix }}%
\providecommand \urlprefix  [0]{URL }%
\providecommand \Eprint [0]{\href }%
\providecommand \doibase [0]{http://dx.doi.org/}%
\providecommand \selectlanguage [0]{\@gobble}%
\providecommand \bibinfo  [0]{\@secondoftwo}%
\providecommand \bibfield  [0]{\@secondoftwo}%
\providecommand \translation [1]{[#1]}%
\providecommand \BibitemOpen [0]{}%
\providecommand \bibitemStop [0]{}%
\providecommand \bibitemNoStop [0]{.\EOS\space}%
\providecommand \EOS [0]{\spacefactor3000\relax}%
\providecommand \BibitemShut  [1]{\csname bibitem#1\endcsname}%
\let\auto@bib@innerbib\@empty
\bibitem [{\citenamefont {Anderson}(1972)}]{Anderson1972}%
  \BibitemOpen
  \bibfield  {author} {\bibinfo {author} {\bibfnamefont {P.~W.}\ \bibnamefont
  {Anderson}},\ }\href {\doibase 10.1126/science.177.4047.393} {\bibfield
  {journal} {\bibinfo  {journal} {Science}\ }\textbf {\bibinfo {volume}
  {177}},\ \bibinfo {pages} {393} (\bibinfo {year} {1972})}\BibitemShut
  {NoStop}%
\bibitem [{\citenamefont {Sachdev}(2011)}]{sachdev_2011}%
  \BibitemOpen
  \bibfield  {author} {\bibinfo {author} {\bibfnamefont {S.}~\bibnamefont
  {Sachdev}},\ }\href {\doibase 10.1017/CBO9780511973765} {\emph {\bibinfo
  {title} {Quantum Phase Transitions}}},\ \bibinfo {edition} {2nd}\ ed.\
  (\bibinfo  {publisher} {Cambridge University Press},\ \bibinfo {address}
  {Cambridge},\ \bibinfo {year} {2011})\BibitemShut {NoStop}%
\bibitem [{\citenamefont {Bohr}\ and\ \citenamefont
  {Mottelson}(1975)}]{Bohr1975}%
  \BibitemOpen
  \bibfield  {author} {\bibinfo {author} {\bibfnamefont {A.}~\bibnamefont
  {Bohr}}\ and\ \bibinfo {author} {\bibfnamefont {B.~R.}\ \bibnamefont
  {Mottelson}},\ }\href@noop {} {\emph {\bibinfo {title} {{Nuclear Structure
  Vols. I and II}}}}\ (\bibinfo  {publisher} {Benjamin},\ \bibinfo {address}
  {New York},\ \bibinfo {year} {1975})\BibitemShut {NoStop}%
\bibitem [{\citenamefont {Grebenev}\ \emph {et~al.}(1998)\citenamefont
  {Grebenev}, \citenamefont {Toennies},\ and\ \citenamefont
  {Vilesov}}]{Grebenev1998}%
  \BibitemOpen
  \bibfield  {author} {\bibinfo {author} {\bibfnamefont {S.}~\bibnamefont
  {Grebenev}}, \bibinfo {author} {\bibfnamefont {J.~P.}\ \bibnamefont
  {Toennies}}, \ and\ \bibinfo {author} {\bibfnamefont {A.~F.}\ \bibnamefont
  {Vilesov}},\ }\href {\doibase 10.1126/science.279.5359.2083} {\bibfield
  {journal} {\bibinfo  {journal} {Science}\ }\textbf {\bibinfo {volume}
  {279}},\ \bibinfo {pages} {2083} (\bibinfo {year} {1998})}\BibitemShut
  {NoStop}%
\bibitem [{\citenamefont {Wenz}\ \emph {et~al.}(2013)\citenamefont {Wenz},
  \citenamefont {Z{\"{u}}rn}, \citenamefont {Murmann}, \citenamefont {Brouzos},
  \citenamefont {Lompe},\ and\ \citenamefont {Jochim}}]{Wenz2013}%
  \BibitemOpen
  \bibfield  {author} {\bibinfo {author} {\bibfnamefont {A.~N.}\ \bibnamefont
  {Wenz}}, \bibinfo {author} {\bibfnamefont {G.}~\bibnamefont {Z{\"{u}}rn}},
  \bibinfo {author} {\bibfnamefont {S.}~\bibnamefont {Murmann}}, \bibinfo
  {author} {\bibfnamefont {I.}~\bibnamefont {Brouzos}}, \bibinfo {author}
  {\bibfnamefont {T.}~\bibnamefont {Lompe}}, \ and\ \bibinfo {author}
  {\bibfnamefont {S.}~\bibnamefont {Jochim}},\ }\href {\doibase
  10.1126/science.1240516} {\bibfield  {journal} {\bibinfo  {journal}
  {Science}\ }\textbf {\bibinfo {volume} {342}},\ \bibinfo {pages} {457}
  (\bibinfo {year} {2013})}\BibitemShut {NoStop}%
\bibitem [{\citenamefont {Launey}(2017)}]{Launey2017}%
  \BibitemOpen
  \bibfield  {author} {\bibinfo {author} {\bibfnamefont {K.~D.}\ \bibnamefont
  {Launey}},\ }\href {\doibase 10.1142/10180} {\emph {\bibinfo {title}
  {Emergent Phenomena in Atomic Nuclei from Large-Scale Modeling}}}\ (\bibinfo
  {publisher} {World Scientific},\ \bibinfo {address} {Singapore},\ \bibinfo
  {year} {2017})\BibitemShut {NoStop}%
\bibitem [{\citenamefont {Bruun}(2014)}]{Bruun2014}%
  \BibitemOpen
  \bibfield  {author} {\bibinfo {author} {\bibfnamefont {G.~M.}\ \bibnamefont
  {Bruun}},\ }\href {\doibase 10.1103/PhysRevA.90.023621} {\bibfield  {journal}
  {\bibinfo  {journal} {Physical Review A - Atomic, Molecular, and Optical
  Physics}\ }\textbf {\bibinfo {volume} {90}},\ \bibinfo {pages} {023621}
  (\bibinfo {year} {2014})}\BibitemShut {NoStop}%
\bibitem [{\citenamefont {Bloch}\ \emph {et~al.}(2008)\citenamefont {Bloch},
  \citenamefont {Dalibard},\ and\ \citenamefont {Zwerger}}]{Bloch2008}%
  \BibitemOpen
  \bibfield  {author} {\bibinfo {author} {\bibfnamefont {I.}~\bibnamefont
  {Bloch}}, \bibinfo {author} {\bibfnamefont {J.}~\bibnamefont {Dalibard}}, \
  and\ \bibinfo {author} {\bibfnamefont {W.}~\bibnamefont {Zwerger}},\ }\href
  {\doibase 10.1103/RevModPhys.80.885} {\bibfield  {journal} {\bibinfo
  {journal} {Reviews of Modern Physics}\ }\textbf {\bibinfo {volume} {80}},\
  \bibinfo {pages} {885} (\bibinfo {year} {2008})}\BibitemShut {NoStop}%
\bibitem [{\citenamefont {Bjerlin}\ \emph {et~al.}(2016)\citenamefont
  {Bjerlin}, \citenamefont {Reimann},\ and\ \citenamefont
  {Bruun}}]{Bjerlin2016}%
  \BibitemOpen
  \bibfield  {author} {\bibinfo {author} {\bibfnamefont {J.}~\bibnamefont
  {Bjerlin}}, \bibinfo {author} {\bibfnamefont {S.~M.}\ \bibnamefont
  {Reimann}}, \ and\ \bibinfo {author} {\bibfnamefont {G.~M.}\ \bibnamefont
  {Bruun}},\ }\href {\doibase 10.1103/PhysRevLett.116.155302} {\bibfield
  {journal} {\bibinfo  {journal} {Physical Review Letters}\ }\textbf {\bibinfo
  {volume} {116}},\ \bibinfo {pages} {155302} (\bibinfo {year}
  {2016})}\BibitemShut {NoStop}%
\bibitem [{som()}]{som}%
  \BibitemOpen
  \href@noop {} {}\bibinfo {note} {See Methods section.}\BibitemShut {Stop}%
\bibitem [{\citenamefont {Sooryakumar}\ and\ \citenamefont
  {Klein}(1980)}]{Sooryakumar1980}%
  \BibitemOpen
  \bibfield  {author} {\bibinfo {author} {\bibfnamefont {R.}~\bibnamefont
  {Sooryakumar}}\ and\ \bibinfo {author} {\bibfnamefont {M.~V.}\ \bibnamefont
  {Klein}},\ }\href {\doibase 10.1103/PhysRevLett.45.660} {\bibfield  {journal}
  {\bibinfo  {journal} {Physical Review Letters}\ }\textbf {\bibinfo {volume}
  {45}},\ \bibinfo {pages} {660} (\bibinfo {year} {1980})}\BibitemShut
  {NoStop}%
\bibitem [{\citenamefont {R{\"{u}}egg}\ \emph {et~al.}(2008)\citenamefont
  {R{\"{u}}egg}, \citenamefont {Normand}, \citenamefont {Matsumoto},
  \citenamefont {Furrer}, \citenamefont {McMorrow}, \citenamefont
  {Kr{\"{a}}mer}, \citenamefont {G{\"{u}}del}, \citenamefont {Gvasaliya},
  \citenamefont {Mutka},\ and\ \citenamefont {Boehm}}]{Ruegg2008}%
  \BibitemOpen
  \bibfield  {author} {\bibinfo {author} {\bibfnamefont {C.}~\bibnamefont
  {R{\"{u}}egg}}, \bibinfo {author} {\bibfnamefont {B.}~\bibnamefont
  {Normand}}, \bibinfo {author} {\bibfnamefont {M.}~\bibnamefont {Matsumoto}},
  \bibinfo {author} {\bibfnamefont {A.}~\bibnamefont {Furrer}}, \bibinfo
  {author} {\bibfnamefont {D.~F.}\ \bibnamefont {McMorrow}}, \bibinfo {author}
  {\bibfnamefont {K.~W.}\ \bibnamefont {Kr{\"{a}}mer}}, \bibinfo {author}
  {\bibfnamefont {H.~U.}\ \bibnamefont {G{\"{u}}del}}, \bibinfo {author}
  {\bibfnamefont {S.~N.}\ \bibnamefont {Gvasaliya}}, \bibinfo {author}
  {\bibfnamefont {H.}~\bibnamefont {Mutka}}, \ and\ \bibinfo {author}
  {\bibfnamefont {M.}~\bibnamefont {Boehm}},\ }\href {\doibase
  10.1103/PhysRevLett.100.205701} {\bibfield  {journal} {\bibinfo  {journal}
  {Physical Review Letters}\ }\textbf {\bibinfo {volume} {100}},\ \bibinfo
  {pages} {205701} (\bibinfo {year} {2008})}\BibitemShut {NoStop}%
\bibitem [{\citenamefont {Bissbort}\ \emph {et~al.}(2011)\citenamefont
  {Bissbort}, \citenamefont {G{\"{o}}tze}, \citenamefont {Li}, \citenamefont
  {Heinze}, \citenamefont {Krauser}, \citenamefont {Weinberg}, \citenamefont
  {Becker}, \citenamefont {Sengstock},\ and\ \citenamefont
  {Hofstetter}}]{Bissbort2011}%
  \BibitemOpen
  \bibfield  {author} {\bibinfo {author} {\bibfnamefont {U.}~\bibnamefont
  {Bissbort}}, \bibinfo {author} {\bibfnamefont {S.}~\bibnamefont
  {G{\"{o}}tze}}, \bibinfo {author} {\bibfnamefont {Y.}~\bibnamefont {Li}},
  \bibinfo {author} {\bibfnamefont {J.}~\bibnamefont {Heinze}}, \bibinfo
  {author} {\bibfnamefont {J.~S.}\ \bibnamefont {Krauser}}, \bibinfo {author}
  {\bibfnamefont {M.}~\bibnamefont {Weinberg}}, \bibinfo {author}
  {\bibfnamefont {C.}~\bibnamefont {Becker}}, \bibinfo {author} {\bibfnamefont
  {K.}~\bibnamefont {Sengstock}}, \ and\ \bibinfo {author} {\bibfnamefont
  {W.}~\bibnamefont {Hofstetter}},\ }\href {\doibase
  10.1103/PhysRevLett.106.205303} {\bibfield  {journal} {\bibinfo  {journal}
  {Physical Review Letters}\ }\textbf {\bibinfo {volume} {106}},\ \bibinfo
  {pages} {205303} (\bibinfo {year} {2011})}\BibitemShut {NoStop}%
\bibitem [{\citenamefont {Endres}\ \emph {et~al.}(2012)\citenamefont {Endres},
  \citenamefont {Fukuhara}, \citenamefont {Pekker}, \citenamefont {Cheneau},
  \citenamefont {Schau{\ss}}, \citenamefont {Gross}, \citenamefont {Demler},
  \citenamefont {Kuhr},\ and\ \citenamefont {Bloch}}]{Endres2012}%
  \BibitemOpen
  \bibfield  {author} {\bibinfo {author} {\bibfnamefont {M.}~\bibnamefont
  {Endres}}, \bibinfo {author} {\bibfnamefont {T.}~\bibnamefont {Fukuhara}},
  \bibinfo {author} {\bibfnamefont {D.}~\bibnamefont {Pekker}}, \bibinfo
  {author} {\bibfnamefont {M.}~\bibnamefont {Cheneau}}, \bibinfo {author}
  {\bibfnamefont {P.}~\bibnamefont {Schau{\ss}}}, \bibinfo {author}
  {\bibfnamefont {C.}~\bibnamefont {Gross}}, \bibinfo {author} {\bibfnamefont
  {E.}~\bibnamefont {Demler}}, \bibinfo {author} {\bibfnamefont
  {S.}~\bibnamefont {Kuhr}}, \ and\ \bibinfo {author} {\bibfnamefont
  {I.}~\bibnamefont {Bloch}},\ }\href {\doibase 10.1038/nature11255} {\bibfield
   {journal} {\bibinfo  {journal} {Nature}\ }\textbf {\bibinfo {volume}
  {487}},\ \bibinfo {pages} {454} (\bibinfo {year} {2012})}\BibitemShut
  {NoStop}%
\bibitem [{\citenamefont {Matsunaga}\ \emph {et~al.}(2013)\citenamefont
  {Matsunaga}, \citenamefont {Hamada}, \citenamefont {Makise}, \citenamefont
  {Uzawa}, \citenamefont {Terai}, \citenamefont {Wang},\ and\ \citenamefont
  {Shimano}}]{Matsunaga2013}%
  \BibitemOpen
  \bibfield  {author} {\bibinfo {author} {\bibfnamefont {R.}~\bibnamefont
  {Matsunaga}}, \bibinfo {author} {\bibfnamefont {Y.~I.}\ \bibnamefont
  {Hamada}}, \bibinfo {author} {\bibfnamefont {K.}~\bibnamefont {Makise}},
  \bibinfo {author} {\bibfnamefont {Y.}~\bibnamefont {Uzawa}}, \bibinfo
  {author} {\bibfnamefont {H.}~\bibnamefont {Terai}}, \bibinfo {author}
  {\bibfnamefont {Z.}~\bibnamefont {Wang}}, \ and\ \bibinfo {author}
  {\bibfnamefont {R.}~\bibnamefont {Shimano}},\ }\href {\doibase
  10.1103/PhysRevLett.111.057002} {\bibfield  {journal} {\bibinfo  {journal}
  {Physical Review Letters}\ }\textbf {\bibinfo {volume} {111}},\ \bibinfo
  {pages} {057002} (\bibinfo {year} {2013})}\BibitemShut {NoStop}%
\bibitem [{\citenamefont {L{\'{e}}onard}\ \emph {et~al.}(2017)\citenamefont
  {L{\'{e}}onard}, \citenamefont {Morales}, \citenamefont {Zupancic},
  \citenamefont {Donner},\ and\ \citenamefont {Esslinger}}]{Leonard2017}%
  \BibitemOpen
  \bibfield  {author} {\bibinfo {author} {\bibfnamefont {J.}~\bibnamefont
  {L{\'{e}}onard}}, \bibinfo {author} {\bibfnamefont {A.}~\bibnamefont
  {Morales}}, \bibinfo {author} {\bibfnamefont {P.}~\bibnamefont {Zupancic}},
  \bibinfo {author} {\bibfnamefont {T.}~\bibnamefont {Donner}}, \ and\ \bibinfo
  {author} {\bibfnamefont {T.}~\bibnamefont {Esslinger}},\ }\href {\doibase
  10.1126/science.aan2608} {\bibfield  {journal} {\bibinfo  {journal}
  {Science}\ }\textbf {\bibinfo {volume} {358}},\ \bibinfo {pages} {1415}
  (\bibinfo {year} {2017})}\BibitemShut {NoStop}%
\bibitem [{\citenamefont {Katsumi}\ \emph {et~al.}(2018)\citenamefont
  {Katsumi}, \citenamefont {Tsuji}, \citenamefont {Hamada}, \citenamefont
  {Matsunaga}, \citenamefont {Schneeloch}, \citenamefont {Zhong}, \citenamefont
  {Gu}, \citenamefont {Aoki}, \citenamefont {Gallais},\ and\ \citenamefont
  {Shimano}}]{Katsumi2018}%
  \BibitemOpen
  \bibfield  {author} {\bibinfo {author} {\bibfnamefont {K.}~\bibnamefont
  {Katsumi}}, \bibinfo {author} {\bibfnamefont {N.}~\bibnamefont {Tsuji}},
  \bibinfo {author} {\bibfnamefont {Y.~I.}\ \bibnamefont {Hamada}}, \bibinfo
  {author} {\bibfnamefont {R.}~\bibnamefont {Matsunaga}}, \bibinfo {author}
  {\bibfnamefont {J.}~\bibnamefont {Schneeloch}}, \bibinfo {author}
  {\bibfnamefont {R.~D.}\ \bibnamefont {Zhong}}, \bibinfo {author}
  {\bibfnamefont {G.~D.}\ \bibnamefont {Gu}}, \bibinfo {author} {\bibfnamefont
  {H.}~\bibnamefont {Aoki}}, \bibinfo {author} {\bibfnamefont {Y.}~\bibnamefont
  {Gallais}}, \ and\ \bibinfo {author} {\bibfnamefont {R.}~\bibnamefont
  {Shimano}},\ }\href {\doibase 10.1103/PhysRevLett.120.117001} {\bibfield
  {journal} {\bibinfo  {journal} {Physical Review Letters}\ }\textbf {\bibinfo
  {volume} {120}},\ \bibinfo {pages} {117001} (\bibinfo {year}
  {2018})}\BibitemShut {NoStop}%
\bibitem [{\citenamefont {Behrle}\ \emph {et~al.}(2018)\citenamefont {Behrle},
  \citenamefont {Harrison}, \citenamefont {Kombe}, \citenamefont {Gao},
  \citenamefont {Link}, \citenamefont {Bernier}, \citenamefont {Kollath},\ and\
  \citenamefont {K{\"{o}}hl}}]{Behrle2018}%
  \BibitemOpen
  \bibfield  {author} {\bibinfo {author} {\bibfnamefont {A.}~\bibnamefont
  {Behrle}}, \bibinfo {author} {\bibfnamefont {T.}~\bibnamefont {Harrison}},
  \bibinfo {author} {\bibfnamefont {J.}~\bibnamefont {Kombe}}, \bibinfo
  {author} {\bibfnamefont {K.}~\bibnamefont {Gao}}, \bibinfo {author}
  {\bibfnamefont {M.}~\bibnamefont {Link}}, \bibinfo {author} {\bibfnamefont
  {J.~S.}\ \bibnamefont {Bernier}}, \bibinfo {author} {\bibfnamefont
  {C.}~\bibnamefont {Kollath}}, \ and\ \bibinfo {author} {\bibfnamefont
  {M.}~\bibnamefont {K{\"{o}}hl}},\ }\href {\doibase 10.1038/s41567-018-0128-6}
  {\bibfield  {journal} {\bibinfo  {journal} {Nature Physics}\ }\textbf
  {\bibinfo {volume} {14}},\ \bibinfo {pages} {781} (\bibinfo {year}
  {2018})}\BibitemShut {NoStop}%
\bibitem [{\citenamefont {Serwane}\ \emph {et~al.}(2011)\citenamefont
  {Serwane}, \citenamefont {Z{\"{u}}rn}, \citenamefont {Lompe}, \citenamefont
  {Ottenstein}, \citenamefont {Wenz},\ and\ \citenamefont
  {Jochim}}]{Serwane2011}%
  \BibitemOpen
  \bibfield  {author} {\bibinfo {author} {\bibfnamefont {F.}~\bibnamefont
  {Serwane}}, \bibinfo {author} {\bibfnamefont {G.}~\bibnamefont {Z{\"{u}}rn}},
  \bibinfo {author} {\bibfnamefont {T.}~\bibnamefont {Lompe}}, \bibinfo
  {author} {\bibfnamefont {T.~B.}\ \bibnamefont {Ottenstein}}, \bibinfo
  {author} {\bibfnamefont {A.~N.}\ \bibnamefont {Wenz}}, \ and\ \bibinfo
  {author} {\bibfnamefont {S.}~\bibnamefont {Jochim}},\ }\href {\doibase
  10.1126/science.1201351} {\bibfield  {journal} {\bibinfo  {journal}
  {Science}\ }\textbf {\bibinfo {volume} {332}},\ \bibinfo {pages} {336}
  (\bibinfo {year} {2011})}\BibitemShut {NoStop}%
\bibitem [{\citenamefont {Kohmoto}\ and\ \citenamefont
  {Takada}(1990)}]{Kohomoto1990}%
  \BibitemOpen
  \bibfield  {author} {\bibinfo {author} {\bibfnamefont {M.}~\bibnamefont
  {Kohmoto}}\ and\ \bibinfo {author} {\bibfnamefont {Y.}~\bibnamefont
  {Takada}},\ }\href {\doibase 10.1143/JPSJ.59.1541} {\bibfield  {journal}
  {\bibinfo  {journal} {Journal of the Physical Society of Japan}\ }\textbf
  {\bibinfo {volume} {59}},\ \bibinfo {pages} {1541} (\bibinfo {year}
  {1990})}\BibitemShut {NoStop}%
\bibitem [{\citenamefont {Nozi{\`{e}}res}\ and\ \citenamefont
  {Pistolesi}(1999)}]{Nozieres1999}%
  \BibitemOpen
  \bibfield  {author} {\bibinfo {author} {\bibfnamefont {P.}~\bibnamefont
  {Nozi{\`{e}}res}}\ and\ \bibinfo {author} {\bibfnamefont {F.}~\bibnamefont
  {Pistolesi}},\ }\href {\doibase 10.1007/s100510050897} {\bibfield  {journal}
  {\bibinfo  {journal} {European Physical Journal B}\ }\textbf {\bibinfo
  {volume} {10}},\ \bibinfo {pages} {649} (\bibinfo {year} {1999})}\BibitemShut
  {NoStop}%
\bibitem [{\citenamefont {Heiselberg}\ and\ \citenamefont
  {Mottelson}(2002)}]{Heiselberg2002a}%
  \BibitemOpen
  \bibfield  {author} {\bibinfo {author} {\bibfnamefont {H.}~\bibnamefont
  {Heiselberg}}\ and\ \bibinfo {author} {\bibfnamefont {B.}~\bibnamefont
  {Mottelson}},\ }\href {\doibase 10.1103/PhysRevLett.88.190401} {\bibfield
  {journal} {\bibinfo  {journal} {Physical Review Letters}\ }\textbf {\bibinfo
  {volume} {88}},\ \bibinfo {pages} {190401} (\bibinfo {year}
  {2002})}\BibitemShut {NoStop}%
\bibitem [{\citenamefont {Bruun}(2002)}]{Bruun2002}%
  \BibitemOpen
  \bibfield  {author} {\bibinfo {author} {\bibfnamefont {G.~M.}\ \bibnamefont
  {Bruun}},\ }\href {\doibase 10.1103/PhysRevLett.89.263002} {\bibfield
  {journal} {\bibinfo  {journal} {Physical Review Letters}\ }\textbf {\bibinfo
  {volume} {89}},\ \bibinfo {pages} {263002} (\bibinfo {year}
  {2002})}\BibitemShut {NoStop}%
\bibitem [{\citenamefont {Rontani}\ \emph {et~al.}(2017)\citenamefont
  {Rontani}, \citenamefont {Eriksson}, \citenamefont {{\AA}berg},\ and\
  \citenamefont {Reimann}}]{Rontani2017}%
  \BibitemOpen
  \bibfield  {author} {\bibinfo {author} {\bibfnamefont {M.}~\bibnamefont
  {Rontani}}, \bibinfo {author} {\bibfnamefont {G.}~\bibnamefont {Eriksson}},
  \bibinfo {author} {\bibfnamefont {S.}~\bibnamefont {{\AA}berg}}, \ and\
  \bibinfo {author} {\bibfnamefont {S.~M.}\ \bibnamefont {Reimann}},\ }\href
  {\doibase 10.1088/1361-6455/aa606a} {\bibfield  {journal} {\bibinfo
  {journal} {Journal of Physics B: Atomic, Molecular and Optical Physics}\
  }\textbf {\bibinfo {volume} {50}},\ \bibinfo {pages} {065301} (\bibinfo
  {year} {2017})}\BibitemShut {NoStop}%
\bibitem [{\citenamefont {Z{\"{u}}rn}\ \emph {et~al.}(2013)\citenamefont
  {Z{\"{u}}rn}, \citenamefont {Lompe}, \citenamefont {Wenz}, \citenamefont
  {Jochim}, \citenamefont {Julienne},\ and\ \citenamefont {Hutson}}]{Zurn2013}%
  \BibitemOpen
  \bibfield  {author} {\bibinfo {author} {\bibfnamefont {G.}~\bibnamefont
  {Z{\"{u}}rn}}, \bibinfo {author} {\bibfnamefont {T.}~\bibnamefont {Lompe}},
  \bibinfo {author} {\bibfnamefont {A.~N.}\ \bibnamefont {Wenz}}, \bibinfo
  {author} {\bibfnamefont {S.}~\bibnamefont {Jochim}}, \bibinfo {author}
  {\bibfnamefont {P.~S.}\ \bibnamefont {Julienne}}, \ and\ \bibinfo {author}
  {\bibfnamefont {J.~M.}\ \bibnamefont {Hutson}},\ }\href {\doibase
  10.1103/PhysRevLett.110.135301} {\bibfield  {journal} {\bibinfo  {journal}
  {Physical Review Letters}\ }\textbf {\bibinfo {volume} {110}},\ \bibinfo
  {pages} {135301} (\bibinfo {year} {2013})}\BibitemShut {NoStop}%
\bibitem [{\citenamefont {Randeria}\ \emph {et~al.}(1990)\citenamefont
  {Randeria}, \citenamefont {Duan},\ and\ \citenamefont
  {Shieh}}]{Randeria1990}%
  \BibitemOpen
  \bibfield  {author} {\bibinfo {author} {\bibfnamefont {M.}~\bibnamefont
  {Randeria}}, \bibinfo {author} {\bibfnamefont {J.~M.}\ \bibnamefont {Duan}},
  \ and\ \bibinfo {author} {\bibfnamefont {L.~Y.}\ \bibnamefont {Shieh}},\
  }\href {\doibase 10.1103/PhysRevB.41.327} {\bibfield  {journal} {\bibinfo
  {journal} {Physical Review B}\ }\textbf {\bibinfo {volume} {41}},\ \bibinfo
  {pages} {327} (\bibinfo {year} {1990})}\BibitemShut {NoStop}%
\bibitem [{\citenamefont {Idziaszek}\ and\ \citenamefont
  {Calarco}(2006)}]{Idziaszek2006}%
  \BibitemOpen
  \bibfield  {author} {\bibinfo {author} {\bibfnamefont {Z.}~\bibnamefont
  {Idziaszek}}\ and\ \bibinfo {author} {\bibfnamefont {T.}~\bibnamefont
  {Calarco}},\ }\href {\doibase 10.1103/PhysRevA.74.022712} {\bibfield
  {journal} {\bibinfo  {journal} {Physical Review A - Atomic, Molecular, and
  Optical Physics}\ }\textbf {\bibinfo {volume} {74}},\ \bibinfo {pages}
  {022712} (\bibinfo {year} {2006})}\BibitemShut {NoStop}%
\bibitem [{\citenamefont {Bruun}\ and\ \citenamefont
  {Mottelson}(2001)}]{Bruun2001}%
  \BibitemOpen
  \bibfield  {author} {\bibinfo {author} {\bibfnamefont {G.~M.}\ \bibnamefont
  {Bruun}}\ and\ \bibinfo {author} {\bibfnamefont {B.~R.}\ \bibnamefont
  {Mottelson}},\ }\href {\doibase 10.1103/PhysRevLett.87.270403} {\bibfield
  {journal} {\bibinfo  {journal} {Physical Review Letters}\ }\textbf {\bibinfo
  {volume} {87}},\ \bibinfo {pages} {270403} (\bibinfo {year}
  {2001})}\BibitemShut {NoStop}%
\bibitem [{\citenamefont {D'Alessio}\ \emph {et~al.}(2016)\citenamefont
  {D'Alessio}, \citenamefont {Kafri}, \citenamefont {Polkovnikov},\ and\
  \citenamefont {Rigol}}]{DAlessio2016}%
  \BibitemOpen
  \bibfield  {author} {\bibinfo {author} {\bibfnamefont {L.}~\bibnamefont
  {D'Alessio}}, \bibinfo {author} {\bibfnamefont {Y.}~\bibnamefont {Kafri}},
  \bibinfo {author} {\bibfnamefont {A.}~\bibnamefont {Polkovnikov}}, \ and\
  \bibinfo {author} {\bibfnamefont {M.}~\bibnamefont {Rigol}},\ }\href
  {\doibase 10.1080/00018732.2016.1198134} {\bibfield  {journal} {\bibinfo
  {journal} {Advances in Physics}\ }\textbf {\bibinfo {volume} {65}},\ \bibinfo
  {pages} {239} (\bibinfo {year} {2016})}\BibitemShut {NoStop}%
\end{thebibliography}%
